\documentclass[english,prl,letterpaper,twocolumn,superscriptaddress,comment]{revtex4}
\usepackage[]{fontenc}
\usepackage[latin9]{inputenc}
\usepackage{amsmath}
\usepackage{amssymb}

\makeatletter
\@ifundefined{textcolor}{}
{%
 \definecolor{BLACK}{gray}{0}
 \definecolor{WHITE}{gray}{1}
 \definecolor{RED}{rgb}{1,0,0}
 \definecolor{GREEN}{rgb}{0,1,0}
 \definecolor{BLUE}{rgb}{0,0,1}
 \definecolor{CYAN}{cmyk}{1,0,0,0}
 \definecolor{MAGENTA}{cmyk}{0,1,0,0}
 \definecolor{YELLOW}{cmyk}{0,0,1,0}
 }

\usepackage{qcircuit}

\makeatother

\newcommand\themap{M}
\newcommand\device{D}
\newcommand\statement{S}

\usepackage{babel}

\begin{document}
\textbf{Verifiable nonlinear quantum evolution implies failure of density matrices to represent proper mixtures}

In a recent Letter, Bennett and coworkers \citep{Bennett2009}
argue that proofs of exotic quantum effects using closed timelike curves (CTC's) based on 
the work of Deutsch \citep{Deutsch1991}, or other nonlinear quantum dynamics, suffer from a 
fallacy that they call the ``linearity trap,'' by which one cannot assume that a classical mixture of 
input states will lead to the same mixture of corresponding output states. We will show that this prescription is inconsistent with the assumption that one can verify the postulated nonlinear evolution.  Specifically, \emph{if some agent can empirically verify the deterministic nonlinear action of a physical device for some set of pure-state inputs, then one cannot generally use density matrices to represent the input for a second agent who is ignorant of the actual pure state prepared.}  Our argument is epistemological and makes no appeal to the detailed physics of CTC's.

The work of \citep{Bennett2009} is motivated by \citep{Brun2009}. As acknowledged in \citep{Bennett2009}, ``there it was shown that for any pair of pure states, $\ket{\phi_0}$ and $\ket{\phi_1}$, there is a CTC-assisted circuit that maps
these to orthogonal states $\ket{0}$ and $\ket{1}$, respectively,'' which we call statement~$\statement$. The two papers agree on~$\statement$, but the authors of \citep{Bennett2009} disagree that it can be used to distinguish nonorthogonal states~\cite{Brun2009}.

If $\statement$ is to be of any empirical consequence, an agent must be able to \emph{verify} it.  If Rob purchases a reusable device~$\device$ (or many such devices, all purported to be the same), which is said to perform the map~$\themap : \ket{\phi_j} \to \ket{j}$, $j \in \{0,1\}$, then Rob will first perform repeated trials to test this.  Rob has two devices~$P_j$ that prepare~$\ket{\phi_j}$, respectively.  During each trial~$n$, he uses $P_{f_n}$ to prepare state~$\ket{\phi_{f_n}}$ and send it into~$\device$, where $f_n$~is an $n$-indexed binary sequence; he measures the output in~$\{\ket{0}, \ket{1} \}$.  Let $C$~label the \emph{procedure} by which $f_n$~is chosen.  Note that many $C$'s may produce the same~$f_n$ (e.g.,~both flipping a fair coin~($C_1$) and choosing to alternate between~0 and~1~($C_2$) might result in~$f_n = \{0,1,0,1,\ldots\}$).

Details of $C$~are irrelevant as long as one such~$C$ exists that would lead Rob to believe~$\device$ deterministically performs~$\themap$ when fed~$\ket{\phi_j}$.  In practice, $C$~must \emph{seem random enough} to Rob (e.g.,~flipping coins, using pseudorandom numbers, choosing by fiat, etc.)\ to convince him of the deterministic, Markovian nature of~$\device$.  If no such~$C$ exists, then Rob concludes that $\device$~does not perform~$\themap$ as advertised.  Asserting~$\statement$ without the possibility of independent verification of~$\themap$ for some~$\device$ is of no empirical consequence, so we assume that such a~$C$ exists for now.

Alice has been watching Rob's verification procedure.  They both note that every time $P_j$~is chosen, measuring $\device$'s output in~$\{\ket{0},\ket{1}\}$ reveals~$\ket{j}$. Sometimes Alice watches the entire procedure; other times she looks away, but all trials succeed, as assumed by verifiability. Rob now asks her to look away while he chooses the device for the next run \emph{through the same process~$C$}. Alice opens her eyes to see the system entering~$\device$.  Alice has looked away before, and her ignorance of the input did not affect the trials' success. By verifiability, Alice's ignorance in this instance similarly cannot influence~$\device$.  By finding $\ket{j}$ as the output, Alice knows that $\ket{\phi_j}$~was the input.

Surprisingly, \citep{Bennett2009} claims that Alice would get no information and that she should have calculated the output of~$\device$ not by looking at~$\themap$ but at how it maps the \emph{density matrix}~$\rho_{A} = p \ket{\phi_0} \bra{\phi_0} + (1-p) \ket{\phi_1} \bra{\phi_1}$ that representing Alice's partial knowledge of the input; she estimates~$p$ from the lab record and other knowledge of~$C$ (e.g.,~coin weighting, etc.).  Since $\rho_A$~is independent of~$f_n$, $\device$'s output would be insensitive to which~$P_j$ was actually chosen. 
But this would remove determinism \emph{even for Rob}, who knew what state he prepared, thus violating verifiability.

The mistake in~\citep{Bennett2009} was to hold too fast to the description of mixtures by density matrices, which was developed for {\it linear} quantum theory.  Decompositions of a density matrix are indistinguishable in ordinary quantum theory, but the proof requires linearity.  With nonlinear evolution, they may be distinguishable, as shown in~\cite{Brun2009} and revisited in~\cite{Ralph2010}.  The authors claim (without justification) that these approaches do not reduce to ordinary quantum theory far from any CTC~\cite{Bennett2009}.  This claim is too strong, since only empirical consistency is required.  Dynamical collapse models are not ruled out by experiments to date and \emph{would} produce preferred ensembles; collapse-free resolutions may also be possible.
%
%
%
%
%

Another possibility is that $\statement$~cannot be verified.  But then no rational agent would believe that any device performs~$\themap$.  In summary, \emph{empirical verifiability of the deterministic, nonlinear action of some physical device for some set of pure-state inputs generally precludes using density matrices to represent \emph{proper mixtures} of such inputs.}  Proper mixtures are based on ignorance of the state \emph{actually} prepared and are a distinct empirical concept from improper mixtures (partial trace of an entangled state).  
In a nonlinear theory, their distinguishability remains an open empirical question.

We thank Jacques Pienaar, Tim Ralph, and Steve Flammia for assistance.
This work was partly funded by an ARC Discovery grant and Postdoctoral Research Fellowship. Research at Perimeter Institute is supported by the Government of Canada through
Industry Canada and by the Province of Ontario through the Ministry
of Research \& Innovation.

Eric G. Cavalcanti$^{\text{1}}$ and Nicolas C. Menicucci$^{\text{2}}$

{\footnotesize 1. Centre for Quantum Dynamics, Griffith University,
Brisbane QLD 4111, Australia}{\footnotesize \par}

{\footnotesize 2. Perimeter Institute for Theoretical Physics, Waterloo,
Ontario N2L 2Y5, Canada}{\footnotesize \par}

\vspace{-2em}

\bibliographystyle{bibstyle_notitle}
\bibliography{CTC_new_arxiv}

\begin{thebibliography}{1}

\vspace{-2em}

\bibitem{Bennett2009}
C.~H. Bennett {\it et~al.}, Phys. Rev. Lett. {\bf 103}, 170502 (2009).

\bibitem{Deutsch1991}
D. Deutsch, Phys. Rev. D {\bf 44}, 3197 (1991).

\bibitem{Brun2009}
T.~A. Brun, J. Harrington, and M.~M. Wilde, Phys. Rev. Lett. {\bf 102}, 210402
  (2009).

\bibitem{Ralph2010}
T.~C. Ralph and C.~R. Myers,  arXiv:1003.1987 [quant-ph] (2010).

\end{thebibliography}

\end{document}